\documentclass[conference, letterpaper]{IEEEtran}

\usepackage{amsmath}
\usepackage{mathtools}
\usepackage{amssymb}
\usepackage[utf8]{inputenc}
\usepackage{xcolor}
\usepackage{graphicx}
\usepackage{svg}
\usepackage{subcaption}
\usepackage[noadjust]{cite}
\usepackage{hyperref}
\usepackage{siunitx}

\usepackage{booktabs}
\usepackage{multirow}
\usepackage{threeparttable}

\usepackage{layouts}

\usepackage{tikz}
\usepackage{tikz-cd}
\usepackage{pgfplots}
\pgfplotsset{compat=1.18}

\usepackage{import}

\title{Compiling Multi-Threshold Operators for Quantized Non-Monotonic and Fine-Granular Activations in Custom-Tailored DNN Accelerators}

\newcommand{\anonymize}[1]{#1}

\author{
\IEEEauthorblockN{
    \anonymize{Christoph Berganski}
}
\IEEEauthorblockA{
    \anonymize{\textit{Computer Engineering Group}}\\
    \anonymize{\textit{Paderborn University}}\\
    \anonymize{Paderborn, Germany}\\
    \anonymize{christoph.berganski@upb.de}
}
\and
\IEEEauthorblockN{
    \anonymize{Felix Jentzsch}
}
\IEEEauthorblockA{
    \anonymize{\textit{Computer Engineering Group}}\\
    \anonymize{\textit{Paderborn University}}\\
    \anonymize{Paderborn, Germany}\\
    \anonymize{felix.jentzsch@upb.de}
}
\and
\IEEEauthorblockN{
    \anonymize{Marco Platzner}
}
\IEEEauthorblockA{
    \anonymize{\textit{Computer Engineering Group}}\\
    \anonymize{\textit{Paderborn University}}\\
    \anonymize{Paderborn, Germany}\\
    \anonymize{platzner@upb.de}
}
}

\begin{document}
    \maketitle
    \begin{abstract}
    Aggressive quantization of neural networks can expose activation functions as bottlenecks for efficient hardware acceleration.
    Implementing quantized activations as multi-threshold operations has been shown to be a promising optimization, allowing for efficient hardware implementation of few-bit activations.
    However, current implementations are limited to monotonically increasing functions, quantization schemes with one-dimensional (per-channel) granularity and a small set of predefined activation-quantizer combinations.
    We formalize the multi-threshold representation to extend it to non-monotonic functions and fine-granular parameters, i.e., multi-dimensional thresholds.
    For automated conversion of quantized function compositions into threshold representation, we propose a systematic approach via algebraic rules integrated into the ONNX-based compiler infrastructure.
    We experimentally validate our solution on a selection of monotonic and non-monotonic functions by extending the binary search based multi-threshold operator that is integrated into the FINN framework for end-to-end synthesis of custom-tailored FPGA accelerators.
    The cost of hardware support for non-monotonicity and fine-granularity is characterized in comparison to an alternative approach utilizing existing operators via graph-level decomposition.
    We compare our method to prior work as well as the approach taken by the hls4ml framework.
    As an outlook, we demonstrate the prospects of using the multi-threshold representation during the training of a neural network as a form of quantization aware training.
\end{abstract}

    \section{Introduction}
\label{sec:introduction}
Quantization has emerged as the key model compression technique for optimizing deep neural network inference~\cite{gholami2022survey,rokh2023comprehensive,10.1145/3527156}. Custom FPGA-based accelerators particularly benefit from aggressive low-bit quantization, typically using 8-bit or lower precisions~\cite{10376654,10.1145/3527156,finn,blott2018finn,fpgaconvnet,duarte2018fast,nn2fpga}. Streaming dataflow architectures exploiting pipelined model parallelism are well suited for such deployments, enabling efficient implementations of flexible quantization schemes. FINN~\cite{finn,blott2018finn, finn-plus} provides an end-to-end compilation flow for models trained and quantized with PyTorch~\cite{Ansel_PyTorch_2_Faster_2024} and Brevitas~\cite{brevitas}, with optimizations expressed through ONNX graph transformations~\cite{Pappalardo:2022nxk,Umuroglu_QONNX_2022}.

Quantization methods can be categorized by when quantization is applied -- post-training quantization (PTQ) or quantization-aware training (QAT) -- and what is quantized, i.e., weights, activations, or both~\cite{gholami2022survey,rokh2023comprehensive,10.1145/3527156}. Early work focused on quantizing weights and multiply-accumulate (MAC) operations in fully connected, convolutional, and attention layers. As these reach increasingly low bit widths, optimizing activations and layer tails, i.e., chains of element-wise operations following a layer, such as normalization, activation, and quantization, becomes critical~\cite{umuroglu2026sira,neseem2024pikelpn,karami2025understanding}. Simultaneously, preserving accuracy requires expressive scale factors, such as floating-point scales, and fine-granular quantization schemes~\cite{krishnamoorthi2018quantizing,nagel2019data,umuroglu2026sira,dai2021vs,yao2022rapq,frantar2023optq}.

At very low bit widths, implementing quantized functions through successive thresholding, also known as multi-threshold functions, has emerged as an effective optimization~\cite{umuroglu2017streamlined,finn,blott2018finn,umuroglu2026sira}. In FINN, multi-threshold functions exactly represent quantized layer tail operations, including non-linearities and floating-point scales, without costly floating-point inference. Here, exact means that no approximation beyond the inherent quantization error is introduced, unlike fixed-point approximations of floating-point values. Enabled by streamlining~\cite{umuroglu2017streamlined,umuroglu2026sira}, scales, biases, and non-linearities are absorbed into threshold values through algebraic transformations, reducing layer tail computations to threshold comparisons. In streaming dataflow architectures, these functions are specialized for each layer tail.

The multi-threshold implementation in FINN is limited to monotonically increasing functions and, in practice, supports only predefined activation–quantizer combinations, i.e, combinations of quantized identity and $\mathrm{Relu}$. However, modern architectures employ diverse activation functions~\cite{DUBEY202292,kunc2024three}, including non-monotonic variants such as $\mathrm{Silu}$~\cite{hendrycks2016gaussian}. Moreover, FINN restricts quantization granularity to the final axis of the activation tensor. While SIRA~\cite{umuroglu2026sira} relaxes some of these constraints, it relies on expensive compile-time simulation of the layer tail.
Consequently, we identify the lack of a systematic and lightweight compilation methodology for translating quantized layer tails into multi-threshold representation and corresponding hardware as a key barrier to a broader adoption.

To extend the flexibility and applicability of the multi-threshold representation, we make the following contributions:
\begin{enumerate}
\item We formalize the multi-threshold representation, including its assumptions, constraints, and principles for efficient hardware acceleration.
\item We extend the representation and hardware architecture to non-monotonic and fine-granular functions.
\item We introduce an ONNX-based compiler integration converting quantized functions into multi-threshold representation and fusing operators through algebraic rules.
\item We present an initial study of directly learning multi-threshold representations as a form of QAT, aligning training-time representation with the target hardware.
\end{enumerate}

The remainder of this paper is structured as follows: Section~\ref{sec:background} introduces background and formalism of the multi-threshold representation, Sections~\ref{sec:non-monotonicity} and~\ref{sec:granularity} extend its applicability by addressing monotonicity and granularity limitations through corresponding hardware implementations. Section~\ref{sec:functions} presents our compiler integration for converting functions into multi-threshold representation. Section~\ref{sec:related-work} discusses related work and contrasts our approach. Section~\ref{sec:learning} explores the prospects of learning multi-threshold representations during training. Section~\ref{sec:conclusion} concludes the paper.

    \section{Background and Formalism}
\label{sec:background}
We define the multi-threshold function $f_{\Theta,\delta}(x)$ with $N$ thresholds $\Theta_{i}$ and associated step direction weights $\delta_{i}$ as
\begin{align}
    f_{\Theta,\delta}(x) = b_{\Theta} + s_{\Theta} \sum_{i=0}^{N-1}\delta_{i} \cdot {\left(x \ge \Theta_{i}\right)},
    \label{eq:multi-threshold}
\end{align}
where $s_{\Theta}$ and $b_{\Theta}$ are output scale and bias parameters.
The thresholds $\Theta_{i}$ can always be sorted in increasing order:
\begin{align}
    \forall i \in \left[1, N-2\right]: \Theta_{i-1}\le\Theta_{i}\le\Theta_{i+1}
    \label{eq:sorted-thresholds}
\end{align}

The representation can be seen as piecewise constant approximation of a function. If $\delta_{i}$ are integers, this relates to a uniform quantizer $\mathbf{Q}$ where $s_{\Theta}$ and $b_{\Theta}$ correspond to quantization scale and bias, respectively:
\begin{align}
    \mathbf{Q}(x) = s_{\Theta}\cdot\text{clip}\left(\left\lfloor\frac{x - b_{\Theta}}{s_{\Theta}}\right\rceil;~0,~2^{\beta}-1\right) + b_{\Theta}
    \label{eq:quantizer}
\end{align}
For static quantization, scale and bias are calibrated to the input range and remain fixed thereafter, and the quantization bitwidth $\beta$ is usually controlled as a hyperparameter.

FINN's multi-threshold implementation~\cite{finn, blott2018finn} has two restrictions. First, as $\delta_{i}$ can take only positive unit steps, the resulting functions are necessarily monotonically increasing:
\begin{align}
    \forall i: \delta_{i} = +1
    \label{eq:monotonicity-constraint}
\end{align}
The relation between quantization bitwidth $\beta$ and the number of thresholds $N$ follows as:
\begin{align}
    \beta = \log_{2}\left(N + 1\right).
    \label{eq:bitwidth}
\end{align}

With this constraint, evaluating the sum in Eq.~\eqref{eq:multi-threshold} can be reduced to locating the input $x$ within the sorted list of thresholds, implicitly setting $\Theta_{-1}=-\infty$ and $\Theta_{N}=+\infty$ to account for inputs outside the threshold range. This can be solved efficiently by binary search~\cite{umuroglu2026sira}.
\begin{align}
    f_{\Theta,\delta}(x) = b_{\Theta} + s_{\Theta}(i+1) \phantom{x}\text{ s.t. } \Theta_{i}\le x < \Theta_{i+1}
    \label{eq:threshold-search}
\end{align}

The second restriction of FINN's multi-threshold implementation concerns the granularity of thresholds: Only per-tensor or per-channel thresholds are permitted, meaning the same set of thresholds is shared by the entire tensor or along slices per axis. 
The shape of the parameter tensor can be expressed as
\begin{align}
    \mathbf{Shape}\left\{\Theta\right\} \subseteq 1 \times 1 \times \dots \times 1 \times C \times N \to C \times N, 
    \label{eq:granularity-constraint}
\end{align}
restricting all leading dimensions to singletons of size $1$, which can be omitted according to tensor broadcasting rules.

    \section{Non-Monotonicity}
\label{sec:non-monotonicity}

\begin{figure*}[h!t]
    \centering
    \includegraphics[scale=1.0]{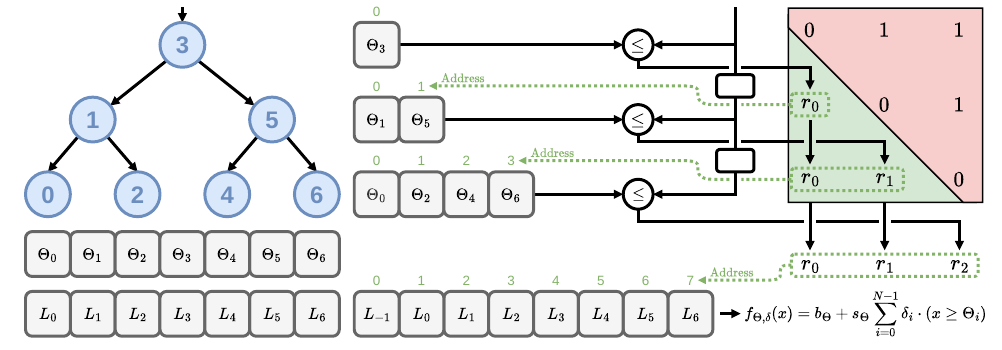}
    \caption{Hardware implementation of binary search based thresholding adapted from~\cite{umuroglu2026sira}. A remapping stage is added to allow for non-monotonic output levels $L_{i}$ according to Eq.~\eqref{eq:non-monotonic-threshold-search}. To locate the input within the list of seven sorted thresholds, three comparators are required, corresponding to the three levels of the search tree shown on the left.}
    \label{fig:schematic:non-monotonicity}
\end{figure*}

In this section we explore the representation of non-monotonic functions by relaxing the constraint on the step weights $\delta_{i}$ to allow for arbitrary integer-valued weights:
\begin{align}
    \forall i: \delta_{i} \in \mathbb{Z}
    \label{eq:integer-steps}
\end{align}

Assuming sorted thresholds, the monotonically increasing output levels of Eq.~\eqref{eq:threshold-search} are replaced by levels $L_{i}$, given as the precomputed cumulative sum over the step weights up until including the largest threshold smaller than the input:
\begin{align}
    \begin{gathered}
        f_{\Theta,\delta}(x) = b_{\Theta} + s_{\Theta}L_{i} \phantom{x}\text{ s.t. } \Theta_{i}\le x < \Theta_{i+1}\\[1ex]
        L_{i} = \sum_{j=0}^{i}\delta_{i},\phantom{x}\text{with }L_{-1} = 0
    \end{gathered}
    \label{eq:non-monotonic-threshold-search}
\end{align}
Where $L_{-1}$ accounts for inputs outside the threshold range. The quantization bitwidth now relates to the levels $L_{i}$ instead of the number of thresholds as
\begin{align}
    \beta^{*} = \log_{2}\left(\max{L_{i}} - \min{L_{i}} + 1\right).
    \label{eq:generalized-bitwidth}
\end{align}
With the monotonicity constraint, Eq.~\eqref{eq:monotonicity-constraint}, this reduces exactly to Eq.~\eqref{eq:bitwidth}, where $\max{L_{i}} = N$ and $\min{L_{i}} = 0$, confirming our generalization to be consistent.

Figure~\ref{fig:schematic:non-monotonicity} sketches our hardware implementation as a remapping of the output, using the output of the threshold search as the address into the precomputed level memory.
We have implemented this in Vitis HLS and integrated it into FINN.

As an alternative to the hardware implementation, we explore a graph-level decomposition for positive and negative unit steps, pulling apart increasing $\delta_{i}=+1$ and decreasing $\delta_{i}=-1$ parts of the sum in Eq.~\eqref{eq:multi-threshold}.
This implements non-monotonic functions as the difference of two multi-threshold functions, each of which is monotonically increasing and operates on a corresponding subset $\Theta^{+}$/$\Theta^{-}$ of the thresholds:
\begin{align}
        f_{\Theta,\delta}(x) = f_{\Theta^{+},\delta=1}(x) - f_{\Theta^{-},\delta=1}(x)
    \label{eq:monotonicity-decomposition}
\end{align}
This reduces the complexity of the thresholding hardware at the cost of additional operators, which besides a second multi-threshold and a subtraction, incurs dataflow overheads.

\begin{figure*}[htbp]
    \centering
    \scalebox{1.0}{\import{figures}{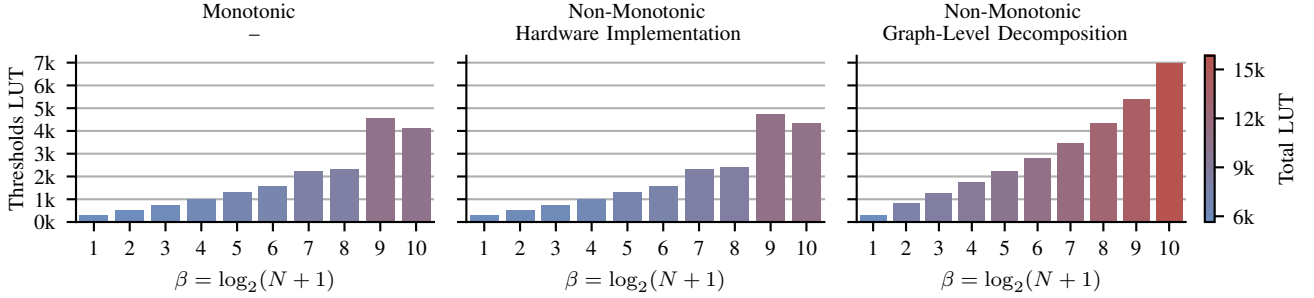}}
    \caption{Comparison of the hardware implementation to the graph-level decomposition in terms of lookup table resource utilization for increasing number of thresholds $N$ when implementing monotonic or non-monotonic functions.}
    \label{fig:results:non-monotonicity}
\end{figure*}
Figure~\ref{fig:results:non-monotonicity} shows results comparing the hardware implementation to the graph-level decomposition in terms of lookup table (LUT) resource utilization.
Our setup is as follows: Thresholds are drawn from a uniform random distribution between $-4$ and $+4$ and non-monotonic step direction weights are generated from a random walk on $\pm1$. Thresholds, as well as the random output scale and bias are single-precision floating-point numbers. In a complete accelerator design, scale and bias would be absorbed into subsequent operations via the streamlining optimization. For our isolated test functions, however, these are implemented and together account for about $5$k LUTs. The shape of the test input is $64 \times 4 \times 4 \times 4$, where the leading $64$ is the batch dimension. A per-tensor set of thresholds and weights is generated. Test functions are exported to ONNX and compiled via FINN+~\cite{finn-plus} with $\mathtt{PE}=4$ parallel processing elements along the final axis, representing a more realistic setting for the FINN framework. The design is synthesized using Vivado 2024.2 targeting the Alveo U55C at \SI{100}{\mega\hertz}. We do not allow DSP or BRAM resources to ensure comparability of results in terms of LUT resources. The design is verified in RTL simulation. We report resource utilization averaged over seven repetitions using varying random seeds.

Non-monotonic functions are more costly to implement than monotonic functions, though the difference is rather small for the direct hardware implementation and on the order of at most a few hundred LUTs. The graph-level decomposition clearly shows the cost of inserting additional operators, where the total LUT utilization is up to $1.5\times$ compared to the hardware implementation. The largest contributions to this increase are the multi-thresholds with $1.6\times$, and dataflow overheads with $3.4\times$ compared to the hardware implementation. Multiplication and addition operations together contribute between $4.8$k and $5.5$k LUTs, with a worst-case increase of about $1.05\times$ compared to the hardware implementation, which is almost negligible.

    \section{Fine Granularity}
\label{sec:granularity}
\begin{figure}[t]
    \centering
    \includegraphics[scale=1.0]{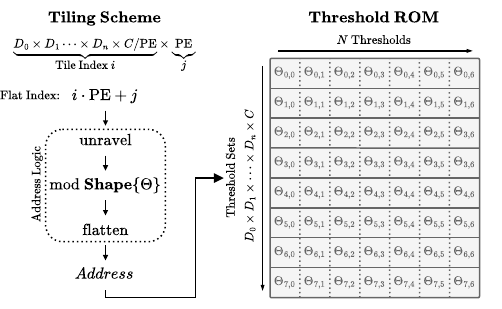}
    \caption{Address logic and layout of the threshold parameter memory. The address logic and the thresholding units are replicated as $\mathtt{PE}$ parallel processing elements resulting in a two-level loop nest $(i,j)$ implemented in Vitis HLS.}
    \label{fig:schematic:granularity}
\end{figure}

In this section we explore the representation of arbitrarily fine-granular threshold parameters.
Fine-granularity is realized via broadcasting of the thresholds to the input shape where the shape of the threshold tensor is a subset of the input shape extended by a trailing dimension of $N$ thresholds:
\begin{align}
    \begin{gathered}
    \mathbf{Shape}\left\{\Theta\right\} \subseteq \mathbf{Shape}\left\{X\right\} \times N\\[1ex]
    \mathbf{Shape}\left\{\Theta\right\} = D_{0} \times D_{1} \times \dots \times D_{n} \times C \times N
    \label{eq:fine-granular}
    \end{gathered}
\end{align}
Address logic maps an index $\mathtt{I} \in \mathbf{Shape}\left\{X\right\}$ to the corresponding index $\mathtt{I}' \in \mathbf{Shape}\left\{\Theta\right\}$ to retrieve the set of thresholds from the parameter memory:
\begin{align}
    \mathtt{I}'_{i} = \mathtt{I}_{i} \mod D_{i}
\end{align}
Memory interfaces and address calculations are realized in terms of flat indices $\mathtt{i}$ and $\mathtt{i}'$, which relate to the multidimensional index via lexicographic ordering.
The inverse, turning a flat index back into a multidimensional index, is referred to as unraveling.
Even though we keep referring to the final axis as the channel axis as this has a special role in FINN, the same also applies to this axis where $C = D_{n+1}$. Figure~\ref{fig:schematic:granularity} sketches the address logic and memory layout of our hardware design.

As an alternative to the hardware implementation, we explore a graph-level decomposition exploiting the property that an arbitrary value can be added to both sides of an equation without changing the outcome.
Following broadcasting semantics, we choose this value such that the resulting thresholds on the right have the desired per-channel granularity:
\begin{align}
    \begin{gathered}
        x \ge \Theta \equiv x + b \ge \Theta', \text{ where } b = \Theta' - \Theta, \text{ and} \\[1ex]
        \mathbf{Shape}\left\{\Theta'\right\} \to C \times N \\[1ex]
        \mathbf{Shape}\left\{b\right\} \to D_{0} \times D_{1} \times \dots \times D_{n} \times C
    \end{gathered}
    \label{eq:granularity-decomposition}
\end{align}
This decomposes thresholds as a fine-granular addition followed by an at most per-channel (last axis) multi-threshold function. However, an exact solution is not always possible as this generally results in reduced parameter volume
\begin{align}
    \left\|\mathbf{Shape}\left\{b\right\}\right\| + \left\|\mathbf{Shape}\left\{\Theta'\right\}\right\| \le \left\|\mathbf{Shape}\left\{\Theta\right\}\right\|,
\end{align}
and we solve the following optimization problem to find the decomposition, which might introduce an approximation error:
\begin{align}
    \Theta', b = \operatorname*{argmin}_{\Theta',b} \|(\Theta - (\Theta' - b)\|_{2}^{2}
\end{align}

\begin{figure*}[htb]
    \centering
    \scalebox{1.0}{\import{figures}{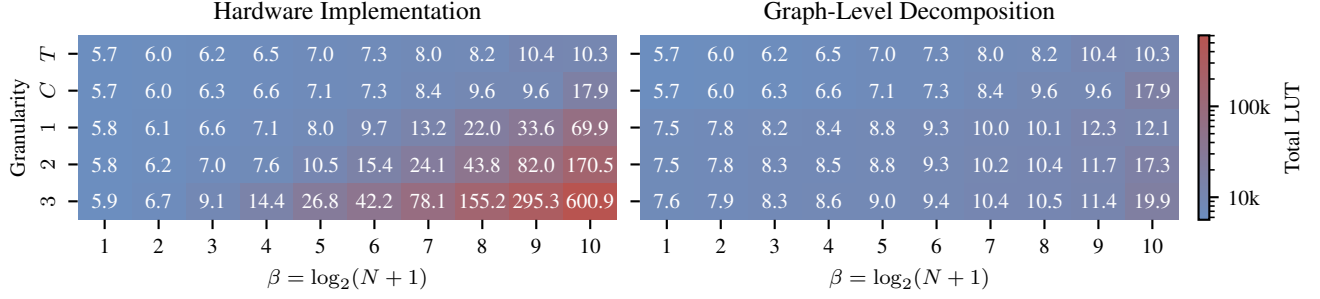}}
    \caption{Comparison of the hardware implementation to the graph-level decomposition  in terms of lookup table resource utilization for increasing number of thresholds $N$ and number of channels covered by parameters. $T$ and $C$ refer to per-tensor and per-channel (last axis) granularity, for which there is  no difference between the two approaches.}
    \label{fig:results:granularity}
\end{figure*}

Figure~\ref{fig:results:granularity} compares the hardware implementation to the graph-level decomposition in terms of LUT resource utilization.
The setup is the same as described in Section~\ref{sec:non-monotonicity}, except for the following: Only monotonically increasing functions are considered, i.e., all step directions are $1$, and instead the granularity is varied by generating random thresholds for varying combinations of axes, which we group by the number of axes covered. Average LUT utilization over seven random seeds per configuration is reported.

Fine-granular parameter tensors are more costly to implement. For per-tensor and per-channel (last axis) granularity there is no difference between the direct hardware implementation and the graph-level decomposition. For fine-granular thresholds at high bitwidths, these results clearly favor the graph-level decomposition with a worst-case increase of about $1.9\times$ from per-tensor to 3-channel granularity. In contrast, the direct hardware implementation, with an increase of about $58\times$, approaches the increase naively expected due to the $64\times$ growth of the parameter tensor.
At this point, the direct hardware implementation becomes impractical, though deviations from this at lower bitwidths (e.g., at $\beta=4$, there is only about a $2.2\times$ difference) hint at automatic logic optimizations of the synthesis tool, which might be possible to exploit at higher bitwidths via more explicit memory partitioning.
At very low bitwidths (i.e., $\beta<4$) and moderate granularities (i.e., 1-2 channels), the direct hardware implementation might be preferable, as decomposition overheads become relevant. 

    \section{Compiling Functions to Thresholds}
\label{sec:functions}
While FINN converts functions into multi-threshold representation via rules for fixed and predefined activation-quantizer patterns and SIRA simulates entire layer tails to extract the steps, we propose absorption rules based on a generalized notion of the inverse $f^{-1}$ of a function:
\begin{align}
    f(x) \ge \Theta_{i} \iff x \ge f^{-1}(\Theta_{i})
    \label{eq:inverse-absorption}
\end{align}

As an example, the generalized inverse of the common $\text{Relu}$ activation function is given as:
\begin{align}
    \text{Relu}^{-1}(x) = \begin{cases}
                              x, & \text{if } x \ge 0 \\ -\infty, & \text{otherwise} \\
    \end{cases}
\end{align}

These inverses need to be predefined for each function of interest, but they are not tied to a particular quantizer nor do they assume a fixed pattern of functions leading up to a quantizer.
No costly simulation of the layer tail is required, keeping conversion runtime and memory footprint low, while reducing the susceptibility to numerical imprecision due to insufficient resolution when sampling the input range.
Range information is used for enumerating the initial thresholds for the rounding function and clipping range of the quantizer,
\begin{align*}
    \mathbf{Q}\left(f(x)\right) =
    b_{\Theta} + s_{\Theta} \sum_{i=0}^{2^{\beta}-1}\left(\frac{f(x) - b_{\Theta}}{s_{\Theta}} \ge \underbrace{\left(i + 0.5 + \Delta_{i}\right)}_{\hat{=}\Theta_{i}} \right),
    \label{eq:proto-thresholds}
\end{align*}
where $\Delta_{i}$ is a round to even correction, which is $\mathbf{Ulp}(i)$ for odd positive and even negative $i$ and zero otherwise. $\mathbf{Ulp}(x)$ is the \emph{unit in the last place}, i.e., the spacing between $x$ and the next representable (floating-point) number.

\begin{figure}[htb]
    \centering
    \includegraphics[scale=1.0]{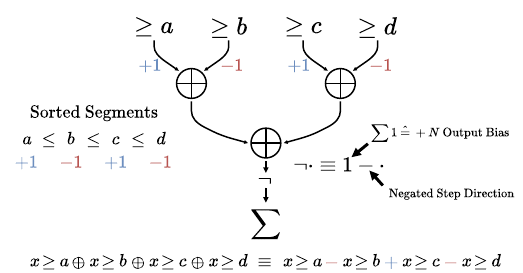}
    \caption{Tree of branching threshold segments: Interpreting this as a parity function, this is equivalent to applying alternating weights $\pm 1$ before summing over the sorted segment contributions. Sorting is always allowed due to commutativity.
    }
    \label{fig:sketch:threshold-segments}
\end{figure}

\begin{figure}[htb]
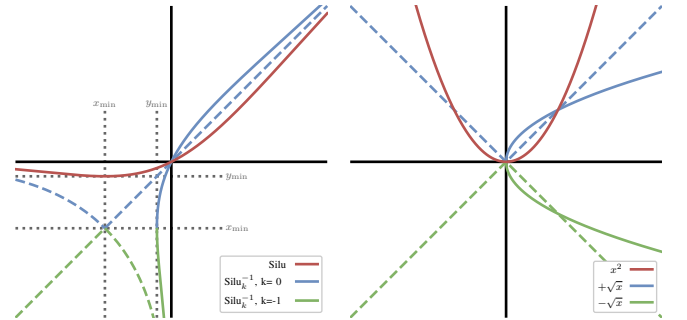

    \centering
    \begin{subfigure}[t]{0.5\columnwidth}
        \scalebox{0.5}{\import{figures}{inverse-silu.pgf}}
    \end{subfigure}%
    \begin{subfigure}[t]{0.5\columnwidth}
        \scalebox{0.5}{\import{figures}{inverse-square.pgf}}
    \end{subfigure}
    \caption{Inverse of non-monotonic functions. Branch points are at the extrema, separating increasing from decreasing segments. Colored dashed lines represent the result of applying the corresponding branch to the output of the function.}
    \label{fig:sketch:inverse}
\end{figure}

Application of Eq.~\eqref{eq:inverse-absorption} to a non-monotonic function with multiple (increasing and decreasing) branches needs to consider all branches as valid solutions, i.e., the set of thresholds branches into corresponding segments.
To give intuition, consider the following example (also depicted in Figure~\ref{fig:sketch:inverse}):
\begin{align*}
    x^{2} \ge a \iff x \ge +\sqrt{a} \lor x \le -\sqrt{a}
\end{align*}
Decreasing branches flip the direction of comparison or, equivalently, invert the step direction, again making use of an $\mathbf{Ulp}$-based correction term:
\begin{align}
    x \ge a \lor x \le b \equiv x \ge a \lor \neg(x \ge b + \mathbf{Ulp}(b))
\end{align}
With $b' = b + \mathbf{Ulp}(b)$, this is rewritten in Xor-form and applied repeatedly to absorb consecutive functions, resulting in a tree of branching threshold segments, which is collapsed to sorted thresholds and step direction weights as shown in Figure~\ref{fig:sketch:threshold-segments}:
\begin{align}
    x \ge a \lor \neg(x \ge b') \equiv & \neg(x \ge b') \oplus \left(x \ge \max\{a, b'\}\right)
\end{align}

We exemplarily derive the generalized inverse of the $\text{Silu}$ activation function~\cite{hendrycks2016gaussian}, which shows a non-monotonic bump to the left of zero.
The $\text{Silu}$ function is defined as follows, where $\sigma(x)$ is the logistic sigmoid function:
\begin{align}
    \text{Silu}(x) = x\cdot\sigma(x)
    \label{eq:silu}
\end{align}

The inverse is expressed in terms of the real-valued principal and secondary branch of the Lambert W function~\cite{corless1996lambert}
\begin{align}
    \text{Silu}^{-1}_{k}(y) = y + W_{k}(ye^{-y}), \text{ where } k\in\left\{0,-1\right\},
    \label{eq:inverse-silu}
\end{align}
which is valid for all $y$ above the global minimum $y_{\min}$ of the $\text{Silu}$ function, which is
\begin{align}
    \begin{aligned}
        x_{\min} &= -1 - W(e^{-1}) \approx -1.28\\
        y_{\min} &= \text{Silu}(x_{\min}) \approx -0.28.
    \end{aligned}
\end{align}
Figure~\ref{fig:sketch:inverse} sketches $\text{Silu}$ with the branches of its inverse, and, similarly, $x^{2}$ with both branches $\pm\sqrt{x}$ of the square root.

We integrate this into the compiler infrastructure as an ONNX function approximating the Lambert W function according to~\cite{iacono2017new} with initial values according to~\cite{LOCZI2022127406}.
Similarly, absorption rules, further generalized inverses, the $\mathbf{Ulp}$, as well as the graph-level decompositions proposed in the previous sections are integrated into the compiler infrastructure\footnote{\url{https://github.com/iksnagreb/onnx-passes}} as ONNX functions and graph transformations.

\begin{figure}[h]
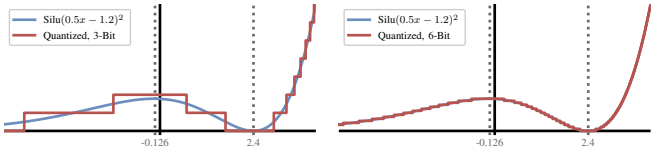

    \centering
    \begin{subfigure}[t]{0.5\columnwidth}
        \scalebox{0.5}{\import{figures}{composition-3.pgf}}
    \end{subfigure}%
    \begin{subfigure}[t]{0.5\columnwidth}
        \scalebox{0.5}{\import{figures}{composition-6.pgf}}
    \end{subfigure}
    \caption{Non-monotonic test function composition at quantization bitwidths $\beta^{*}=3$ and $\beta^{*}=6$. The function has two increasing and one decreasing segment with branch points at approximately $(-0.126,0.078)$ and exactly $(2.4,0.0)$.}
    \label{fig:sketch:silu-squared}
\end{figure}
\begin{figure*}[ht!]
    \centering
    \scalebox{1.0}{\import{figures}{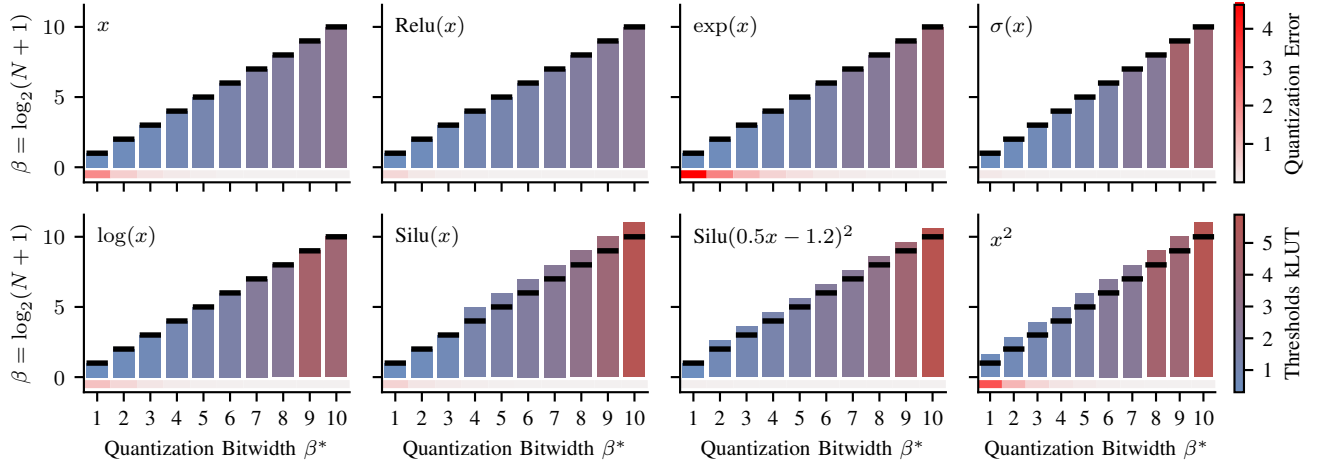}}
    \caption{Comparison of functions implemented as multi-threshold functions in terms of lookup table resource utilization over increasing quantization bitwidth $\beta^{*}$.
    The thresholding bitwidth $\beta$ results from calibrating the quantizer to the function at the bitwidth $\beta^{*}$.
    For monotonic functions $\beta = \beta^{*}$, as indicated by the black horizontal bars. Non-monotonic functions might revisit the same output level, resulting in $\beta > \beta^{*}$.
    Functions are ordered from top left to bottom right by average LUTs per output level. The quantization error (mean absolute error) decreases with the quantization bitwidth as indicated below each bar.}
    \label{fig:results:functions}
\end{figure*}

Figure~\ref{fig:results:functions} shows the resource utilization for various monotonic and non-monotonic functions approximated with increasing quantization bitwidth $\beta^{*}$.
We demonstrate operator fusion via absorption by implementing the following non-monotonic function composition as a single multi-threshold function:
\begin{align}
    \mathbf{Q}\left(f(x)\right) = \mathbf{Q}\left(\text{Silu}(0.5 x - 1.2)^{2}\right)
    \label{eq:example-composition}
\end{align}
This function at two different quantization bitwidths is shown in Figure~\ref{fig:sketch:silu-squared}.
The experimental setup is similar to Section~\ref{sec:non-monotonicity}, except for the following: As there is no randomness in these functions, a single result is reported without averaging.
Thresholds result from calibrating and converting a uniform quantizer on the functions evaluated over the $-4$ to $+4$ range ($\log$ is evaluated over $+0.01$ to $+8.01$).
The number of thresholds now is a dependent variable and instead the quantization bitwidth $\beta^{*}$, i.e., number output levels, is controlled.

Non-monotonic functions are on average more costly to implement. This can be primarily attributed to requiring more thresholds at the same quantization bitwidth, i.e., usually $\beta > \beta^{*}$. Besides the raw number of thresholds, the concrete values have an influence on the resource utilization as well: functions with more uniform threshold distribution, such as the identity or $\text{Relu}$, require fewer LUTs at the same quantization bitwidth. Ordering functions by their average LUT utilization per output level matches an intuitive notion of non-uniformity.

    \section{Related Work}
\label{sec:related-work}
In this section, we first qualitatively compare our work to prior approaches for compiling multi-threshold operators and then provide a quantitative comparison to hls4ml. 

\subsection{Qualitative Comparison}
Table~\ref{tab:related:finn} presents the qualitative comparison. Originally~\cite{finn,blott2018finn}, FINN provides a small set of fixed activation-quantizer patterns which are converted to a direct implementation of the summation realized in HLS. SIRA~\cite{umuroglu2026sira} proposes an automated compilation which works out of the box for any operator supported by ONNX Runtime and thus does not require operator-specific extensions. Alongside, a more efficient binary search based implementation of the multi-threshold operator is introduced, which is available as an RTL backend in FINN. To avoid the high compilation effort of subgraph evaluation and search, we propose a semi-automatic compilation, where primitives are implemented manually but are combined automatically by following function composition along the layer tail to support more complex activation-quantizer patterns. We extend frontend and backend to support non-monotonic and arbitrarily fine-granular functions. Furthermore, it is possible to use our frontend to target the SIRA backend via graph-level decomposition.
\begin{table}[htb]
    \centering
    \begin{tabular}{c c c c}
    \toprule
         & \textbf{Original} & \textbf{SIRA} & \multirow{2}{*}{\textbf{Ours}} \\
         & \cite{finn,blott2018finn} & \cite{umuroglu2026sira} & \\ \midrule
         \textbf{Compilation} & & & \\
          \multirow{2}{*}{Approach} & \multirow{2}{*}{\shortstack{Manual\\\emph{Fixed Patterns}}} & \multirow{2}{*}{\shortstack{Automatic\\\emph{Search-Based}}} & \multirow{2}{*}{\shortstack{Semi-Automatic\\\emph{Fusing Primitives}}} \\
          &&& \\[1ex]
         Effort & Low & High\textsuperscript{1} & Low \\ \midrule
         \textbf{Support} &&& \\
         \multirow{2}{*}{\shortstack{Monotonicity}} & \multirow{2}{*}{Monotonic} & \multirow{2}{*}{Monotonic} & \multirow{2}{*}{\shortstack{Monotonic\\Non-Monotonic}} \\
         &&& \\[1ex]
         Granularity & Per-Channel & Per-Channel & Arbitrary \\ \midrule
         \textbf{Backend} &&& \\
         \multirow{2}{*}{\shortstack{Algorithm}} & \multirow{2}{*}{Summation} & \multirow{2}{*}{Binary Search} & \multirow{2}{*}{Binary Search} \\
         &&& \\
         Implementation & HLS & RTL & HLS/RTL\textsuperscript{2} \\ \midrule
         \multirow{2}{*}{\textbf{Extensibility}} & \multirow{2}{*}{\shortstack{Add Converter\\per Pattern}} & \multirow{2}{*}{--\textsuperscript{3}} & \multirow{2}{*}{\shortstack{Add Inverse\\per Primitive}} \\
         &&& \\
     \bottomrule
    \end{tabular}
    \begin{itemize}
        \scriptsize
        \item[1.] Search space resolution trades off effort vs. compilation error
        \item[2.] SIRA backend with decomposition of non-monotonic/fine-granular functions
        \item[3.] Subgraph evaluation works out of the box for standard operators
    \end{itemize}
    \caption{Layer tail compilation approaches in FINN}
    \label{tab:related:finn}
\end{table}

\subsection{Quantitative Comparison}
The hls4ml framework~\cite{duarte2018fast,fastml_hls4ml,10.1145/3801979} employs a lookup table approach to implement a variety of activation functions. The input range is uniformly split into sections and function output values are tabulated to be retrieved at runtime. hls4ml does not offer operator fusion, such as the streamlining optimization~\cite{umuroglu2017streamlined,umuroglu2026sira} in FINN. Rather, hls4ml implements each operator in a layer tail as a standalone hardware operator, which we refer to as composite approach.

\begin{figure*}[h!h]
    \centering
    \scalebox{1.0}{\import{figures}{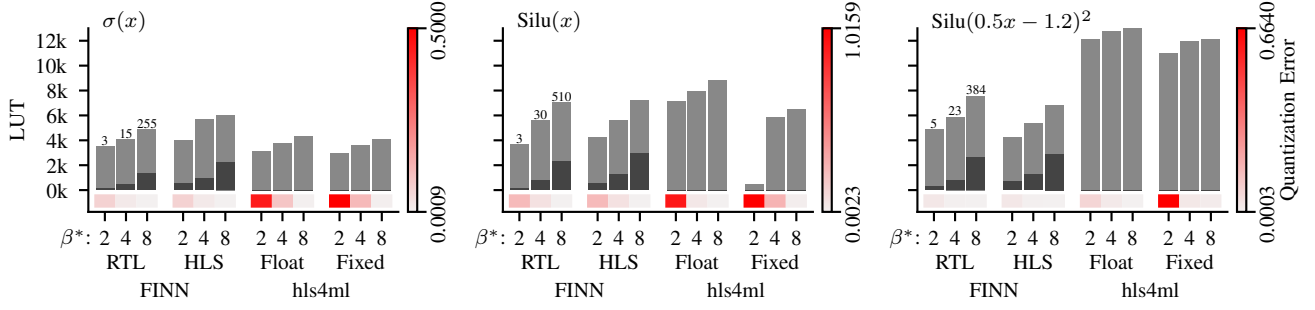}}
    \caption{Comparison of functions implemented as multi-threshold operators in FINN and a composite lookup-table based approach in hls4ml in terms of LUT count over quantization bitwidth $\beta^{*}$. For FINN, \emph{HLS} refers to our binary-search backend (Section~\ref{sec:non-monotonicity}) and \emph{RTL} refers to the backend introduced by SIRA, which requires graph-level decomposition for the non-monotonic functions.
    Shaded in dark grey is the portion of LUTs used by the multi-threshold operators, which is a more realistic estimate of the resource impact within a streamlined accelerator, where de-quantization occurs only once at the end of the model graph, not after each operation.
    For hls4ml we set the number of lookup table entries to the same number of thresholds required by our approach -- this number is annotated on top of the first group of bars per function. %
    The selected lookup table data type is either floating-point or $\beta^{*}$ bit fixed-point.
    The two functions on the right are not directly supported as a lookup table in hls4ml and are realized as a composition of a $\sigma(x)$ lookup table and standalone additions and multiplications.
    }
    \label{fig:related:hls4ml}
\end{figure*}

To compare our approach to hls4ml as well as FINN extended by the SIRA~\cite{umuroglu2026sira} RTL backend, we use a setup similar to Section~\ref{sec:functions},  configuring hls4ml to generate a pipelined design with $\texttt{II}=1$ and a reuse factor matching our $\texttt{PE}=4$ parallelism.
The results are presented in Figure~\ref{fig:related:hls4ml}.
For simpler activation functions which are supported out of the box, hls4ml can implement the function with comparatively fewer resources, but at the cost of higher quantization errors, especially for few-bit quantization.
We attribute this to the uniform sampling of the input range, whereas the non-uniformity of the multi-thresholds allows to better allocate the intervals to steeper regions of the function, which can also be seen in Figure~\ref{fig:sketch:silu-squared}.
For more complex function compositions, the resource costs of the multi-threshold approach grow only moderately due to operator fusion.
By streamlining a model graph, which delays the de-quantization scale and bias to the very end of the graph (dark portion of the bars in Figure~\ref{fig:related:hls4ml}), this has the potential to implement the entire quantized activation data-path in just a few hundred LUTs (at, for example, 4-bit quantization).
In contrast, the composite approach taken by hls4ml becomes largely unattractive due to the stark increase in resource utilization.
Comparing the two FINN implementations, the SIRA RTL backend is generally more efficient in terms of thresholds per LUT.
However, as SIRA does not directly support implementation of non-monotonic functions, decomposition overheads start to dominate for more complex non-monotonic function compositions, making our HLS backend more attractive.

    \section{Learning Thresholds}
\label{sec:learning}
As an outlook, we explore the prospects of learning quantized activation functions directly in multi-threshold representation.
This offers an option for training aggressively quantized neural networks from scratch, which avoids having to find the ideal activation function at the desired bitwidth, and, by making the quantization thresholds flexible, brings back some of the representational capacity usually lost due to low-bitwidth quantization.
Thus, we can interpret learning multi-threshold representations as a form of quantization-aware training. Learning of activation functions has been proposed before, usually by expressing activation functions as learnable polynomials~\cite{goyal2019learning,goyal2020improved}, rational functions~\cite{fang2023transformers} or splines~\cite{liu2024kan}, whereas we propose a linear combination of step functions.

We relax the non-differentiable comparison as a smooth sigmoid function $\sigma$ centered at each threshold.
The sharpness of the step can be controlled by a factor $\alpha$, which for $\alpha \gg 1$ closely approximates the comparison.
For $\|w_{i}\| \approx 1$ the step weights $\delta_{i} \in \left\{-1,+1\right\}$ can be made differentiable via the straight through estimator trick: $\delta_{i} = \mathtt{sign}(w_{i}) \approx w_{i}$.
Thus, to compute gradients during the backward pass, we treat the multi-threshold function as follows:
\begin{align}
    f_{\Theta,\delta}(x) \approx b_{\Theta} + s_{\Theta} \sum_{i=0}^{N-1}{w_{i} \cdot \sigma\left(\alpha \cdot \left(x - \Theta_{i}\right)\right)}
    \label{eq:relaxation}
\end{align}

This yields the following partial derivatives with respect to the function input $x$ and the function parameters $\Theta_{j}$ and $w_{j}$, where $\sigma_{i} = \sigma\left(\alpha\left(x - \Theta_{i}\right)\right)$:
\begin{align}
    \frac{\partial f_{\Theta,\delta}}{\partial x} \approx s_{\Theta} \sum_{i=0}^{N-1}\alpha w_{i} \sigma_{i} \left(1 - \sigma_{i}\right)
\end{align}
\begin{align}
    \frac{\partial f_{\Theta,\delta}}{\partial \Theta_{j}} \approx - \alpha w_{j} \sigma_{j} \left(1 - \sigma_{j}\right)
\end{align}
\begin{align}
    \frac{\partial f_{\Theta,\delta}}{\partial w_{j}} \approx \sigma_{j}
\end{align}

\begin{figure*}[htp!]
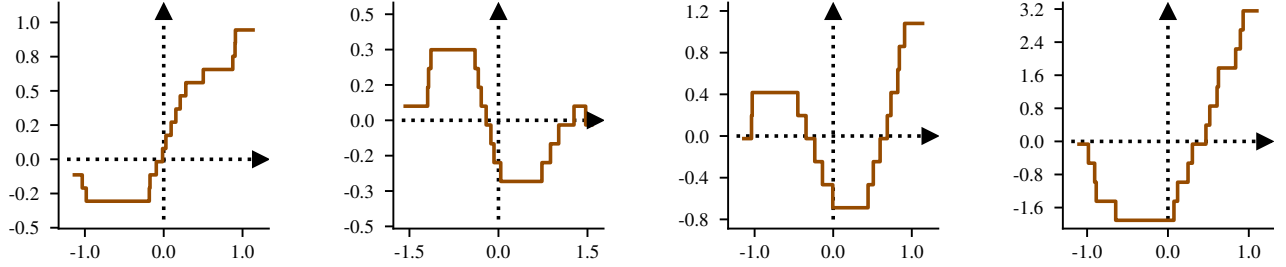

    \centering
    \begin{subfigure}[t]{0.5\columnwidth}
        \centering
        \scalebox{0.9}{\import{figures}{thresholds-000-000.pgf}}
    \end{subfigure}%
    \begin{subfigure}[t]{0.5\columnwidth}
        \centering
        \scalebox{0.9}{\import{figures}{thresholds-001-000.pgf}}
    \end{subfigure}%
    \begin{subfigure}[t]{0.5\columnwidth}
        \centering
        \scalebox{0.9}{\import{figures}{thresholds-002-000.pgf}}
    \end{subfigure}%
    \begin{subfigure}[t]{0.5\columnwidth}
        \centering
        \scalebox{0.9}{\import{figures}{thresholds-003-000.pgf}}
    \end{subfigure}
    \caption{Non-monotonic per-tensor multi-threshold functions learned by the four activation layers of LeNet for $N=15$ steps.
    }
    \label{fig:results:learend-functions}
\end{figure*}
\begin{figure}[ht!]
    \centering
    \scalebox{0.5}{\input{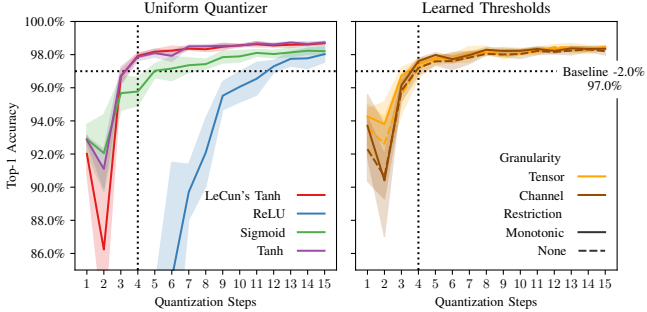}}
    \caption{Comparison of learned multi-threshold functions to hand-selected quantized activation functions for MNIST classification with LeNet. 
    Standard deviation in shaded.
    }
    \label{fig:results:learning-mnist}
\end{figure}
We implement this alongside the sharp forward pass in PyTorch and demonstrate the effectiveness by QAT experiments over a range of quantization steps for a simple MNIST classification task with LeNet~\cite{726791}.
We train these models for 40 epochs minimizing the cross entropy loss using the Adam optimizer~\cite{kingma2015adam} with learning rate $0.001$ and no weight decay.
A non-quantized baseline is trained reaching about \SI{99}{\percent} classification accuracy.
We allow up to 2 percentage points accuracy drop with quantization.
The results in comparison to hand-selected activation functions are shown in Figure~\ref{fig:results:learning-mnist}, showing comparable results.
In both settings, the model can be quantized down to 4 steps corresponding to approximately \SI{2.32}{\bit}, before the accuracy starts to break down completely.
The accuracy increase observed at 1-bit quantization needs to be investigated in future work, but seems unrelated to the learned thresholds as it is observed with the uniform quantizer as well.
While the results for the uniform quantizer heavily depend on the choice of activation function, the results suggest the learnable multi-threshold functions to be more robust in this low-bitwidth quantization regime.
Figure~\ref{fig:results:learend-functions} exemplarily shows functions learned for an instance of the non-monotonic per-tensor setting. The learned functions seem plausible in the sense, that they are centered at about zero and cover a range which is typical for neural network activation functions.

    \section{Conclusion}
\label{sec:conclusion}
We have extended the multi-threshold representation of quantized functions to non-monotonic functions with fine-granular parameters by extending the binary search based multi-threshold operator integrated into the FINN framework and an alternative graph-level decomposition approach.
The proposed solutions have been evaluated in terms of resource utilization on an FPGA to characterize the cost of non-monotonicity and fine-granularity.
A systematic and light-weight compilation approach of quantized layer tails to multi-threshold representation has been proposed to facilitate end-to-end synthesis of custom-tailored FPGA accelerators.
We have successfully demonstrated this on a selection of monotonic and non-monotonic functions.
Furthermore, we have proposed directly learning multi-threshold representations as a form of QAT, which we have successfully demonstrated on a simple MNIST classifier.
Future work will extend function support by implementing the necessary inverse functions and further investigate quality and performance implications of the newly gained options by using them with more demanding neural network inference workloads.

    \section*{Acknowledgments}
\anonymize{This work has been funded by the Ministry of Culture and Science of the State of North Rhine-Westphalia under the grant no NW21-059D (SAIL) and by the German Federal Ministry for the Environment, Nature Conservation, Nuclear Safety and Consumer Protection under the grant no 67KI32004A (eki). \\

We gratefully acknowledge the computing time provided to us on the high-performance computers Noctua 2 and Otus at the NHR Center PC2.
}

    \bibliographystyle{IEEEtran}
\bibliography{references}

@software{finn-plus,
  author       = {Jentzsch, Felix and Berganski, Christoph and Jungemann, Linus and Wintermann, Bjarne},
  title        = {{FINN+}},
  year         = {2026},
  publisher    = {Zenodo},
  doi          = {10.5281/zenodo.19100351},
  url          = {https://github.com/eki-project/finn-plus}
}

@inproceedings{finn,
    author = {Umuroglu, Yaman and Fraser, Nicholas J. and Gambardella, Giulio and Blott, Michaela and Leong, Philip and Jahre, Magnus and Vissers, Kees},
    title = {FINN: A Framework for Fast, Scalable Binarized Neural Network Inference},
    booktitle = {Proceedings of the 2017 ACM/SIGDA International Symposium on Field-Programmable Gate Arrays},
    series = {FPGA '17},
    year = {2017},
    pages = {65--74},
    publisher = {ACM}
}

@article{blott2018finn,
    title = {FINN-R: An end-to-end deep-learning framework for fast exploration of quantized neural networks},
    author = {Blott, Michaela and Preu{\ss}er, Thomas B and Fraser, Nicholas J and Gambardella, Giulio and O’brien, Kenneth and Umuroglu, Yaman and Leeser, Miriam and Vissers, Kees},
    journal = {ACM Transactions on Reconfigurable Technology and Systems (TRETS)},
    volume = {11},
    number = {3},
    pages = {1--23},
    year = {2018},
    publisher = {ACM New York, NY, USA}
}

@article{umuroglu2017streamlined,
    title = {Streamlined deployment for quantized neural networks},
    author = {Umuroglu, Yaman and Jahre, Magnus},
    journal = {arXiv preprint arXiv:1709.04060},
    year = {2017}
}

@article{umuroglu2026sira,
author = {Umuroglu, Yaman and Berganski, Christoph and Jentzsch, Felix and Danilowicz, Michal and Kryjak, Tomasz and Bezaitis, Charalampos and Sj\"{a}lander, Magnus and Colbert, Ian and Preusser, Thomas and Petri-Koenig, Jakoba and Blott, Michaela},
title = {SIRA: Scaled-Integer Range Analysis for Optimizing FPGA Dataflow Neural Network Accelerators},
year = {2026},
issue_date = {June 2026},
publisher = {Association for Computing Machinery},
address = {New York, NY, USA},
volume = {19},
number = {2},
issn = {1936-7406},
url = {https://doi.org/10.1145/3807510},
doi = {10.1145/3807510},
journal = {ACM Transactions on Reconfigurable Technology and Systems},
month = may,
articleno = {23},
numpages = {33}
}

@inproceedings{Ansel_PyTorch_2_Faster_2024,
author = {Ansel, Jason and Yang, Edward and He, Horace and Gimelshein, Natalia and Jain, Animesh and Voznesensky, Michael and Bao, Bin and Bell, Peter and Berard, David and Burovski, Evgeni and Chauhan, Geeta and Chourdia, Anjali and Constable, Will and Desmaison, Alban and DeVito, Zachary and Ellison, Elias and Feng, Will and Gong, Jiong and Gschwind, Michael and Hirsh, Brian and Huang, Sherlock and Kalambarkar, Kshiteej and Kirsch, Laurent and Lazos, Michael and Lezcano, Mario and Liang, Yanbo and Liang, Jason and Lu, Yinghai and Luk, CK and Maher, Bert and Pan, Yunjie and Puhrsch, Christian and Reso, Matthias and Saroufim, Mark and Siraichi, Marcos Yukio and Suk, Helen and Suo, Michael and Tillet, Phil and Wang, Eikan and Wang, Xiaodong and Wen, William and Zhang, Shunting and Zhao, Xu and Zhou, Keren and Zou, Richard and Mathews, Ajit and Chanan, Gregory and Wu, Peng and Chintala, Soumith},
booktitle = {29th ACM International Conference on Architectural Support for Programming Languages and Operating Systems, Volume 2 (ASPLOS '24)},
doi = {10.1145/3620665.3640366},
month = apr,
publisher = {ACM},
title = {{PyTorch 2: Faster Machine Learning Through Dynamic Python Bytecode Transformation and Graph Compilation}},
url = {https://docs.pytorch.org/assets/pytorch2-2.pdf},
year = {2024}
}

@software{brevitas,
  author       = {Franco, Giuseppe and Pappalardo, Alessandro and Fraser, Nicholas J},
  title        = {Xilinx/brevitas},
  year         = {2025},
  publisher    = {Zenodo},
  doi          = {10.5281/zenodo.3333552},
  url          = {https://doi.org/10.5281/zenodo.3333552}
}

@inproceedings{Pappalardo:2022nxk,
    author = "Pappalardo, Alessandro and Umuroglu, Yaman and Blott, Michaela and Mitrevski, Jovan and Hawks, Ben and Tran, Nhan and Loncar, Vladimir and Summers, Sioni and Borras, Hendrik and Muhizi, Jules and Trahms, Matthew and Hsu, Shih-Chieh Hsu and Hauck, Scott and Duarte, Javier",
    title = "{QONNX: Representing Arbitrary-Precision Quantized Neural Networks}",
    booktitle = "{4th Workshop on Accelerated Machine Learning (AccML) at HiPEAC 2022 Conference}",
    eprint = "2206.07527",
    archivePrefix = "arXiv",
    primaryClass = "cs.LG",
    reportNumber = "FERMILAB-CONF-22-471-SCD",
    month = "6",
    year = "2022",
    url = "https://accml.dcs.gla.ac.uk/papers/2022/4thAccML_paper_1(12).pdf"
}

@software{Umuroglu_QONNX_2022,
author = {Umuroglu, Yaman and Borras, Hendrik and Loncar, Vladimir and Summers, Sioni and Duarte, Javier},
doi = {10.5281/zenodo.7622236},
license = {Apache-2.0},
month = jun,
title = {{QONNX}},
url = {https://github.com/fastmachinelearning/qonnx},
version = {0.1},
year = {2022}
}

@article{duarte2018fast,
  title={Fast inference of deep neural networks in FPGAs for particle physics},
  author={Duarte, Javier and Han, Song and Harris, Philip and Jindariani, Sergo and Kreinar, Edward and Kreis, Benjamin and Ngadiuba, Jennifer and Pierini, Maurizio and Rivera, Ryan and Tran, Nhan and others},
  journal={Journal of instrumentation},
  volume={13},
  number={07},
  pages={P07027},
  year={2018},
  publisher={IOP Publishing}
}

@software{fastml_hls4ml,
  author       = {{FastML Team}},
  title        = {fastmachinelearning/hls4ml},
  year         = 2026,
  publisher    = {Zenodo},
  version      = {v1.3.0},
  doi          = {10.5281/zenodo.1201549},
  url          = {https://github.com/fastmachinelearning/hls4ml}
}

@article{10.1145/3801979,
author = {Schulte, Jan-Frederik and Ramhorst, Benjamin and Sun, Chang and Mitrevski, Jovan and Ghielmetti, Nicol\`{o} and Lupi, Enrico and Danopoulos, Dimitrios and Lon\v{c}ar, Vladimir and Duarte, Javier and Burnette, David and Laatu, Lauri and Tzelepis, Stylianos and Axiotis, Konstantinos and Berthet, Quentin and Wang, Haoyan and White, Paul and Demirsoy, Suleyman and Colombo, Marco and Aarrestad, Thea Klaeboe and Summers, Sioni and Pierini, Maurizio and Di Guglielmo, Giuseppe and Ngadiuba, Jennifer and Campos, Javier and Hawks, Benjamin and Gandrakota, Abhijith and Fahim, Farah and Tran, Nhan and Constantinides, George A and Que, Zhiqiang and Luk, Wayne and Tapper, Alexander and Hoang, Duc and Paladino, Noah and Harris, Philip C. and Lai, Bo-Cheng and Valentin, Manuel and Forelli, Ryan and Ogrenci, Seda and Gerlach, Lino and Brooks Flynn, Rian and Liu, Mia and Diaz, Daniel and Koda, Elham E. and Quinnan, Melissa and Solares, Russell Marroquin and Parajuli, Santosh and Neubauer, Mark S. and Herwig, Christian and Tsoi, Ho Fung and Rankin, Dylan and Hsu, Shih-Chieh and Hauck, Scott},
title = {hls4ml: A Flexible, Open Source Platform for Deep Learning Acceleration on Reconfigurable Hardware},
year = {2026},
issue_date = {June 2026},
publisher = {Association for Computing Machinery},
address = {New York, NY, USA},
volume = {19},
number = {2},
issn = {1936-7406},
url = {https://doi.org/10.1145/3801979},
doi = {10.1145/3801979},
journal = {ACM Trans. Reconfigurable Technol. Syst.},
month = may,
articleno = {19},
numpages = {35}
}

@ARTICLE{nn2fpga,
    author={Bosio, Roberto and Minnella, Filippo and Urso, Teodoro and Casu, Mario R. and Lavagno, Luciano and Lazarescu, Mihai T. and Pasini, Paolo},
    journal={IEEE Transactions on Computer-Aided Design of Integrated Circuits and Systems}, 
    title={{NN2FPGA}: Optimizing CNN Inference on FPGAs With Binary Integer Programming}, 
    year={2024},
    volume={},
    number={},
    pages={1-1},
    doi={10.1109/TCAD.2024.3507570}
}

@INPROCEEDINGS{fpgaconvnet,
  author={Venieris, Stylianos I. and Bouganis, Christos-Savvas},
  booktitle={2016 IEEE 24th Annual International Symposium on Field-Programmable Custom Computing Machines (FCCM)}, 
  title={fpgaConvNet: A Framework for Mapping Convolutional Neural Networks on FPGAs}, 
  year={2016},
  volume={},
  number={},
  pages={40-47},
  doi={10.1109/FCCM.2016.22}
}

@incollection{gholami2022survey,
  title={A survey of quantization methods for efficient neural network inference},
  author={Gholami, Amir and Kim, Sehoon and Dong, Zhen and Yao, Zhewei and Mahoney, Michael W and Keutzer, Kurt},
  booktitle={Low-power computer vision},
  pages={291--326},
  year={2022},
  publisher={Chapman and Hall/CRC}
}

@article{rokh2023comprehensive,
  title={A comprehensive survey on model quantization for deep neural networks in image classification},
  author={Rokh, Babak and Azarpeyvand, Ali and Khanteymoori, Alireza},
  journal={ACM Transactions on Intelligent Systems and Technology},
  volume={14},
  number={6},
  pages={1--50},
  year={2023},
  publisher={ACM New York, NY}
}

@article{10.1145/3527156,
author = {Armeniakos, Giorgos and Zervakis, Georgios and Soudris, Dimitrios and Henkel, J\"{o}rg},
title = {Hardware Approximate Techniques for Deep Neural Network Accelerators: A Survey},
year = {2022},
issue_date = {April 2023},
publisher = {Association for Computing Machinery},
address = {New York, NY, USA},
volume = {55},
number = {4},
issn = {0360-0300},
url = {https://doi.org/10.1145/3527156},
doi = {10.1145/3527156},
journal = {ACM Comput. Surv.},
month = nov,
articleno = {83},
numpages = {36}
}

@INPROCEEDINGS{10376654,
  author={Colbert, Ian and Pappalardo, Alessandro and Petri-Koenig, Jakoba},
  booktitle={2023 IEEE/CVF International Conference on Computer Vision (ICCV)}, 
  title={A2Q: Accumulator-Aware Quantization with Guaranteed Overflow Avoidance}, 
  year={2023},
  volume={},
  number={},
  pages={16943-16952},
  doi={10.1109/ICCV51070.2023.01558}}

@inproceedings{neseem2024pikelpn,
  title={{PikeLPN}: Mitigating Overlooked Inefficiencies of Low-Precision Neural Networks},
  author={Neseem, Marina and McCullough, Conor and Hsin, Randy and Leichner, Chas and Li, Shan and Chong, In Suk and Howard, Andrew and Lew, Lukasz and Reda, Sherief and Rautio, Ville-Mikko and others},
  booktitle={Proceedings of the IEEE/CVF Conference on Computer Vision and Pattern Recognition},
  pages={15996--16005},
  year={2024}
}

@inproceedings{karami2025understanding,
  title={Understanding the Performance Horizon of the Latest ML Workloads with NonGEMM Workloads},
  author={Karami, Rachid and Kao, Sheng-Chun and Kwon, Hyoukjun},
  booktitle={2025 IEEE International Symposium on Performance Analysis of Systems and Software (ISPASS)},
  pages={1--14},
  year={2025},
  organization={IEEE}
}

@article{dai2021vs,
  title={{VS-Quant}: Per-vector scaled quantization for accurate low-precision neural network inference},
  author={Dai, Steve and Venkatesan, Rangha and Ren, Mark and Zimmer, Brian and Dally, William and Khailany, Brucek},
  journal={Proceedings of Machine Learning and Systems},
  volume={3},
  pages={873--884},
  year={2021}
}

@article{yao2022rapq,
  title={{RAPQ}: Rescuing accuracy for power-of-two low-bit post-training quantization},
  author={Yao, Hongyi and Li, Pu and Cao, Jian and Liu, Xiangcheng and Xie, Chenying and Wang, Bingzhang},
  journal={arXiv preprint arXiv:2204.12322},
  year={2022}
}

@inproceedings{frantar2023optq,
  title={{OPTQ}: Accurate post-training quantization for generative pre-trained transformers},
  author={Frantar, Elias and Ashkboos, Saleh and Hoefler, Torsten and Alistarh, Dan-Adrian},
  booktitle={11th International Conference on Learning Representations},
  year={2023}
}

@inproceedings{nagel2019data,
  title={Data-free quantization through weight equalization and bias correction},
  author={Nagel, Markus and Baalen, Mart van and Blankevoort, Tijmen and Welling, Max},
  booktitle={Proceedings of the IEEE/CVF international conference on computer vision},
  pages={1325--1334},
  year={2019}
}

@article{krishnamoorthi2018quantizing,
  title={Quantizing deep convolutional networks for efficient inference: A whitepaper},
  author={Krishnamoorthi, Raghuraman},
  journal={arXiv preprint arXiv:1806.08342},
  year={2018}
}

@article{corless1996lambert,
  title={On the Lambert W function},
  author={Corless, Robert M and Gonnet, Gaston H and Hare, David EG and Jeffrey, David J and Knuth, Donald E},
  journal={Advances in Computational mathematics},
  volume={5},
  number={1},
  pages={329--359},
  year={1996},
  publisher={Springer}
}

@article{iacono2017new,
  title={New approximations to the principal real-valued branch of the Lambert W-function},
  author={Iacono, Roberto and Boyd, John P},
  journal={Advances in Computational Mathematics},
  volume={43},
  number={6},
  pages={1403--1436},
  year={2017},
  publisher={Springer}
}

@article{LOCZI2022127406,
title = {Guaranteed- and high-precision evaluation of the Lambert W function},
journal = {Applied Mathematics and Computation},
volume = {433},
pages = {127406},
year = {2022},
issn = {0096-3003},
doi = {https://doi.org/10.1016/j.amc.2022.127406},
url = {https://www.sciencedirect.com/science/article/pii/S0096300322004805},
author = {Lajos Lóczi}
}

@article{hendrycks2016gaussian,
  title={Gaussian Error Linear Units (Gelus)},
  author={Hendrycks, D},
  journal={arXiv preprint arXiv:1606.08415},
  year={2016}
}

@article{DUBEY202292,
title = {Activation functions in deep learning: A comprehensive survey and benchmark},
journal = {Neurocomputing},
volume = {503},
pages = {92-108},
year = {2022},
issn = {0925-2312},
doi = {https://doi.org/10.1016/j.neucom.2022.06.111},
url = {https://www.sciencedirect.com/science/article/pii/S0925231222008426},
author = {Shiv Ram Dubey and Satish Kumar Singh and Bidyut Baran Chaudhuri}
}

@article{kunc2024three,
  title={Three decades of activations: A comprehensive survey of 400 activation functions for neural networks},
  author={Kunc, Vladim{\'\i}r and Kl{\'e}ma, Ji{\v{r}}{\'\i}},
  journal={arXiv preprint arXiv:2402.09092},
  year={2024}
}

@article{goyal2019learning,
  title={Learning activation functions: A new paradigm for understanding neural networks},
  author={Goyal, Mohit and Goyal, Rajan and Lall, Brejesh},
  journal={arXiv preprint arXiv:1906.09529},
  year={2019}
}

@inproceedings{goyal2020improved,
  title={Improved polynomial neural networks with normalised activations},
  author={Goyal, Mohit and Goyal, Rajan and Lall, Brejesh},
  booktitle={2020 International Joint Conference on Neural Networks (IJCNN)},
  pages={1--8},
  year={2020},
  organization={IEEE}
}

@inproceedings{fang2023transformers,
  title={Transformers with learnable activation functions},
  author={Fang, Haishuo and Lee, Ji-Ung and Moosavi, Nafise Sadat and Gurevych, Iryna},
  booktitle={Findings of the Association for Computational Linguistics: EACL 2023},
  pages={2382--2398},
  year={2023}
}

@article{liu2024kan,
  title={Kan: Kolmogorov-arnold networks},
  author={Liu, Ziming and Wang, Yixuan and Vaidya, Sachin and Ruehle, Fabian and Halverson, James and Solja{\v{c}}i{\'c}, Marin and Hou, Thomas Y and Tegmark, Max},
  journal={arXiv preprint arXiv:2404.19756},
  year={2024}
}

@ARTICLE{726791,
  author={Lecun, Y. and Bottou, L. and Bengio, Y. and Haffner, P.},
  journal={Proceedings of the IEEE}, 
  title={Gradient-based learning applied to document recognition}, 
  year={1998},
  volume={86},
  number={11},
  pages={2278-2324},
  doi={10.1109/5.726791}
}

@inproceedings{kingma2015adam,
  title={Adam: {A} Method for Stochastic Optimization},
  author={Kingma, Diederik P. and Ba, Jimmy},
  booktitle={International Conference on Learning Representations ({ICLR})},
  year={2015}
}

\end{document}